# Demonstration of MIMO-DFS over 100km of unamplified SSMF Link using Active Laser Drift Stabilization and Optimized Probing Codes

**Rajiv Boddeda[1], Christian Dorize[1], Pierre Brochard[2], Haïk Mardoyan[1], Carina Castineiras[1] and Jérémie Renaudier[1]**

*(1) Nokia Bell Labs, 12 rue Jean Bart, 91300 Massy, France   (2) Silentsys, 10 rue Xavier Bichat, 72000, Le Mans, France*
*rajiv.boddeda@nokia-bell-labs.com*

**Abstract:** We estimate the laser frequency noise impact on coherent sensing using Distributed Fiber Sensing model. By stabilizing the laser in the estimated frequency zone, we demonstrate reduced noise floor over 100km using optimized probing codes. © 2025 The Author(s)

## 1. Introduction

Reusing the installed optical telecom infrastructure for sensing environmental vibrations is gaining interest from both civil society [1] and telecom operators [2]. As standard distributed fiber sensing (DFS) techniques are mainly targeted to sense dedicated fiber links and are usually limited to relatively short links, we introduced [3] a sensing approach derived from coherent optical transmission where a laser source is phase modulated by digital codes jointly probing two orthogonal polarization axes using a polarization diversity coherent receiver at the reception. This multiple input and multiple output (MIMO)-DFS approach brings two major benefits: 1) polarization fading is mitigated, leading to substantial sensitivity gain compared with interrogation over a single polarization channel [4]; 2) polarization diversity digital coding interrogation does not induce any penalty to transmission performed on adjacent WDM channels and so can be safely introduced in live telecom networks [5]. This paper aims at further enhancing sensitivity while increasing optical reach in DFS systems without resorting to Raman amplification [6] nor chirped pulse techniques capable of detecting a particular region in the fiber [7]. We consider the interaction between a time-spreading interrogation of a fiber link and the laser instabilities encountered during the backscattered channel estimation. We first simulate the impact of the laser frequency noise power spectral density (PSD) on the DFS sensitivity along an SSMF, comparing a true laser response to a standard Lorentzian model approach [8]. We observe that the rise of the frequency noise in the spectral region below 10 kHz strongly degrades the DFS sensitivity. We then introduce in our MIMO-DFS setup a system that compensates for the low frequency variations of the sensing laser source and show the DFS sensitivity gain. We finally report the sensitivity measured in our lab along a 123 km SSMF link, highlighting the ability of MIMO-DFS used jointly with a stabilized laser source to capture environmental noise within a 382 Hz mechanical bandwidth with a 1 m spatial resolution.

## 2. Quantifying impact of laser frequency noise on DFS sensitivity

Sensitivity and reach performances are limited by two independent effects in practical DFS implementations: additive noise from the optoelectronic components and coherence loss of the laser source. The former effect is magnified by the ultra-weak backscattered intensity level in a SMF along with the speckle noise that randomly occurs along a fiber link. However, coherent optical mixing used in modern DFS implementations partly compensates for it; and digital processing techniques exist to alleviate SNR issues [6,9]. The latter effect, due to instabilities of laser source central frequency, remains the major source of performance loss. This is particularly true in recent DFS systems aiming at long distance sensing and that modulate the laser light with a signal spread in time, as frequency chirps [7] or digital codes [3,6], instead of older intensity-based pulse interrogation methods. DFS performs consecutive backscattered channel estimations with a period $T_p$ at least equal to the round-trip-time $T_r$ in the fiber to sense. With time-spreading interrogation, the modulation of the light per channel estimation is usually fixed to $T_p$ for SNR concerns. The convolution of the light modulation with the backscattered channel thus spreads over a support equal to $T_{conv}= T_p +T_r$, whereas it is equal to $T_r$ with pulse interrogation. Therefore, this interrogation technique is likely to be sensitive to laser frequency instability since any slow fluctuation of the carrier during the period $T_{conv}$ corrupts the channel estimation. We simulate this effect using our DFS simulation environment by including laser frequency noise on the modulated probing code at Tx and on the coherent mixer at Rx [10].

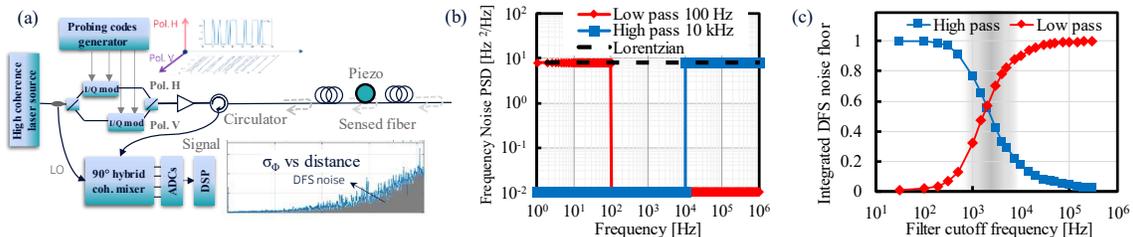

Fig. 1: a) MIMO-DFS setup for sensing the fiber, b) Principle for laser frequency noise filtering from an 8Hz$^2$/Hz Lorentzian example case, c) Integrated noise floor as a function of cut-off frequencies, DFS frequency zone of impact is shaded in grey.

We simulate a PDM-BPSK probing code modulation [9] to capture the backscattered response of an emulated 50 km length SMF ($T_r = 0.488$ ms). The code is made of $2^{17}$ symbols, the modulation rate is 50 Mbaud, yielding $T_p =2.62$ ms code duration and 2 m gauge length. 500 codes are consecutively transmitted, emulating 1.3 s SMF measurement in a noiseless environment. In Fig. 1a, we show the MIMO DFS setup with polarization division multiplexed IQ modulator where the retro-reflected signal from the sensed fiber is separated using an optical circulator and detected using coherent receiver. The optical phase variation is estimated locally at each segment position as a standard deviation value expressed in radian, leading to a distance dependent noise floor as displayed in the right part of Fig. 1a.

We now introduce a method to determine the spectral zone of laser frequency noise which mostly contributes to the DFS noise floor. Our criterion is the noise floor integrated along the overall sensed fiber distance, which comes down to the surface below the curve shown on the right of Fig. 1a. We perform a series of DFS noise floor simulations, partly filtering out the laser frequency noise thanks to a high-pass and a low-pass digital filters (raised cosine, 0.1 roll-off factor) with variable cut-off frequency, see Fig. 1b with a low pass filter at 100 Hz and high pass filter at 10 kHz. We finally estimate the integrated noise floor achieved with each laser frequency noise filtered version, for which noise floor integration is fixed to unit norm. Fig. 1c shows the frequency noise impact as a function of the cut-off frequency value for the high-pass (blue curve) and the low-pass (red curve) filters in the case of a 25 Hz linewidth Lorentzian model laser source, so with a constant frequency noise power spectral density (PSD) value of 25/pi=8 $Hz^2$/Hz. We observe that 50% of the DFS integrated noise floor originates from the laser frequency noise bandwidth [800-6k] Hz, centered at 2 kHz. These values are related to the term $T_{conv} = 311$ ms in this simulation (1/311e$^{-3}$ = 322 Hz corresponding sine wave frequency).

Free-running lasers exhibit frequency noise which differ from the ideal Lorentzian case: their sensitivity to the environment (vibrations, acoustics, temperature, current supply…) induces a so-called "technical noise" that further increases the white noise level when going towards low frequencies. To better reflect the impact of a true laser source, we replace the previous Lorentzian case by a high-quality sensing dedicated Teraxion NLL laser source. Its frequency noise PSD measured in our lab and displayed in red color in Fig. 2b also shows a frequency noise of about 8 $Hz^2$/Hz in its Lorentzian region beyond 10 kHz, but an expected increase below 10 kHz. This response was introduced in our DFS simulation environment [10] to emulate its impact on DFS sensitivity. We found that the frequency noise impact is further transposed towards low frequency compared with the Lorentzian case, with a maximal impact at 550 Hz instead of 2 kHz and 50% of the DFS noise floor induced by frequency noise in the range [100-2.5k] Hz.

This study shows for the first time to the best of our knowledge, that DFS noise floor is only induced by slow variations, of the light source around its central frequency, and the spectral zone of impact can be precisely determined from $T_r$ in the fiber to sense, the probing length $T_p$, and the frequency noise PSD response of the used laser source, so delivering engineering rules to further enhance sensitivity of DFS systems.

**3. Mitigating laser frequency noise through external control**

As all laser sources show an increase of their frequency noise towards low frequency, and especially in the spectral region of maximal impact for DFS system sensitivity, hence lowering the laser frequency noise in the spectral region of interest is crucial; in other worlds, introducing a stabilization of our laser source slow fluctuations. The key element for laser frequency stabilization is the optical frequency discriminator (OFD) that gives in real time the laser frequency fluctuation information. It can be a reference laser, a Fabry-Pérot cavity, a wavemeter, an atomic transition or absorption. For our experiment, as we aim to reach low frequency noise roughly between 10 Hz and 10 kHz, we need a laser source that has both low white frequency noise level and high-bandwidth frequency modulation capability (typ. >100 kHz); and we need a high quality and high bandwidth OFD. Based on the building blocks from Silentsys, we conceived a compact OFD system operating in standard telecom bands for sensing applications, as shown in Fig. 2a.

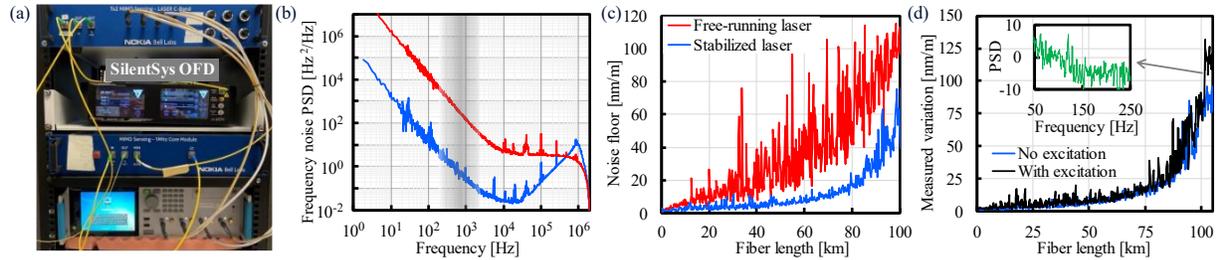

Fig. 2: a) MIMO sensing setup including the OFD system, b) PSD of frequency noise for laser source without and with external control, c) Noise floor achieved experimentally along the first 100 km of an SMF without and with external control, d) Experimental detection of a 120 Hz (see inset) sinewave excitation at 101 km

The laser frequency fluctuations are converted to voltage fluctuations by the OFD which are in turn applied to a servo controller whose output is connected to the frequency modulation input port of the Teraxion NLL laser. To achieve a good frequency locking, the servo controller is optimized by locking on the OFD resonance to reduce as low as possible the noise in the bandwidth of interest. Fig. 2b displays the frequency noise PSD of the laser source measured prior and after frequency stabilization with the OFD. It shows that the noise is reduced by a factor of $10^3$ for frequencies below 10 kHz. The slight bump observed at 1 MHz is an effect of the stabilization control loop but is transparent for DFS noise floor according to our simulation results, as it is far from the laser frequency spectral region of influence.

## 4. Experimental results

We now experimentally validate the impact of the laser frequency noise reduction thanks to the OFD based control loop. We choose a 123 km SMF spool to reflect the maximal span length found in telecom networks, since DFS is basically planned to be implemented on a per span basis. We use advanced PDM-MPSK probing code design derived from [9], achieving a shorter code length $T_p$ that approaches the physical limit of the fiber for sensing, hence allowing to relax the frequency stabilization restrictions on a shorter timescale and therefore towards a higher frequency zone in PSD, see shaded region in Fig. 2b. The code is made of $2^{17}$ symbols that modulate the laser light with 1 dBm output power at 100 Mbaud, leading to a 1 m native gauge length, and the permanent repetition of the code allows to capture mechanical events over a 382 Hz bandwidth. With $T_r = 0.98$ ms and $T_p = 1.31$ ms, the term $T_{conv} = 2.3$ ms is in the same range as in the above simulation case, so this DFS experiment is subject to a similar spectral region of laser fluctuations. We estimate the DFS phase standard deviation as a function of distance and convert it into the measurand elongation of fiber per meter [11]. Fig. 2c displays the noise floor along the first 100 kms of the SMF measured successively without and with active laser noise control. As expected from simulation results, we observe a drastic decrease of the DFS noise floor. We complement the study by provoking a 120 Hz sine wave excitation at 101 km from fiber start, where a piezo actuator induces a 180 nε peak to peak strain over 1.3 m. The black curve in Fig. 2d shows the strong increase above the noise floor (in blue) at the excitation position. We verified by spectral analysis the clear emergence of the 120 Hz component 10 dB above the PSD noise floor at 101 kms as shown in Fig. 2d inset.

**Conclusion**

We showed using simulations how the laser frequency noise impacts the sensitivity of the DFS system using the signal from the Rayleigh backscattering. By targeting the specific spectral range where frequency noise significantly affects the DFS noise floor, the laser frequency was stabilized, resulting in a greater than twofold reduction in the DFS noise floor. This improvement enabled the detection of low energy events beyond 100 kilometers. This opens the door towards highly sensitive DFS systems use over existing fiber infrastructures with fiber spans up to 100 kms.